\documentclass[aps,prl,amssymb,showpacs]{revtex4}
\usepackage{graphicx}

\def \ff {\frac}
\begin{document}
\title{Pattern Selection in a Phase Field Model for Directional
 Solidification}
\author{R. N. Costa Filho$^{1}$, J. M. Kosterlitz$^{2}$, and Enzo
Granato$^{3}$}
\affiliation{$^1$Departamento de F\'{i}sica,
Universidade Federal do  Cear\'a, Caixa Postal 6030, Campus do Pici,
60451-970 Fortaleza, Cear\'a, Brasil}
\affiliation{$^2$Department of Physics, Brown University, Providence,
 RI02912, USA}
\affiliation{$^3$Laborat\'{o}rio Associado de Sensores e Materiais,
 Instituto Nacional de Pesquisas Espaciais,
12201-970 S\~{a}o Jos\'{e} dos Campos, SP Brasil}

\begin{abstract}
A symmetric phase field model is used to study wavelength selection in
 two dimensions. We study the problem in a finite
system using a two-pronged approach. First we construct an action and,
 minimizing this, we obtain the most probable
configuration of the system, which we identify with the selected
 stationary state. The minimization is constrained by
the stationary solutions of stochastic evolution equations and is done
 numerically. Secondly, additional support for
this selected state is obtained from straightforward simulations of
 the dynamics from a variety of initial states.
\end{abstract}

\maketitle
\section{Introduction}
The problem of pattern selection is to predict which pattern, out of a
number of possible stationary patterns, will be selected under given
experimental conditions. This problem has been of  great interest for
many years in many fields such as biology, chemistry, engineering and
physics  \cite{CrossHohenberg} where a large variety of reproducible
patterns can be formed under various external conditions which drive
the system  away from thermal and mechanical equilibrium. Important
examples of pattern selection are directional solidification  and
eutectic growth where the interface between the ordered and disordered
phases can take a periodic cellular  pattern. From the experimental
point of view, the interface seems to have a well defined periodicity
in both directed one phase  solidification
\cite{bechhoefer87,billia87,flesselles91} and in directed eutectic
solidification \cite{jacksonhunt66,ginibre97}. On the theoretical
side, there is much disagreement with some researchers  claiming that
a unique wavelength is selected
\cite{karma86,kerszberg1,kerszberg2,kerszberg3,kurtze96} while others
say that the wavelength of the final pattern is accidental
\cite{dombre87,amar88}. A large scale  simulation on a noisy Swift
Hohenberg equation \cite{swifthohenberg} which models fluid convection
near onset has  been done \cite{garcia93} to study the selection of
periodic patterns. For this system, a potential ${\cal  F}_{SH}$
exists and minimizing this gives the unique selected stationary
pattern for any initial state,  exactly as for a system approaching
equilibrium.

\par In this paper, we use a numerical and a theoretical approach to
argue that in the presence of noise, there is a wavelength selection
for the cellular pattern formation in the directional solidification
problem, see Fig. 1. More precisely, we use a phase field model with
additive stochastic noise developed for growth processes such as
single phase directional solidification driven out of equilibrium by
an external moving temperature gradient
\cite{grossmann93,drolet94,drolet00}. This causes invasion of a
disordered by an ordered phase with non potential dynamics. Since we
are unable to solve the problem analytically, we use numerical
computations which limit us to rather small system sizes. However, we
believe that this is sufficient to demonstrate our conjecture of a
unique selected state. In addition, we construct an
action and,  minimizing this, we obtain the most probable
configuration of the system, which we identify with the selected
stationary state. The minimization is constrained by the stationary
solutions of stochastic evolution equations which are calculated
numerically.
\begin{figure}[h]
\includegraphics[width=5.5cm]{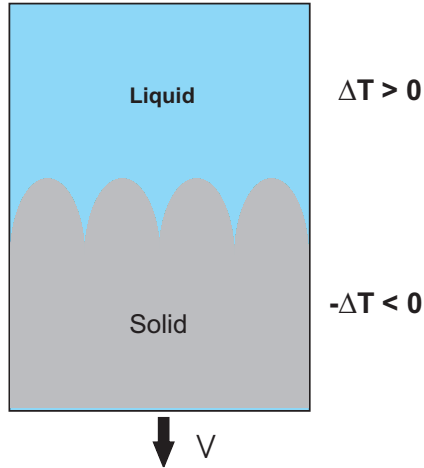} \caption{A schematic
representation of the cellular pattern in  directional
solidification. $\Delta T=(T-T_m)/T_m$ where $T_m$ is the melting
temperature of the pure solid.}
 \label{fig1} \end{figure}


\section{Phase-field model}

To carry out our simulations, we use a phase-field model for
directional solidification based on two continuum fields to
describe the phases of the system. These fields vary slowly in the
bulk region and rapidly near the solid-liquid interface. The
equilibrium system is described by a free energy functional ${\cal
F}(c,\psi)$ \cite{grossmann93} where
\begin{eqnarray}
\label{eq:free1}
{\cal F}(c,\psi)&=&\int d{\bf r}
\left[\frac{K_\psi}{2}(\nabla\psi)^2+\frac{K_{c}}{2}(\nabla
 c)^{2}+f(c,\psi)\right] \cr
f(c,\psi)&=&-\ff{1}{2}\psi^2+\ff{1}{4}\psi^4+\Delta Th(\psi)
+\ff{1}{2}\gamma\Delta\psi
 \left(c+\frac{h(\psi)}{\Delta\psi}\right)^{2} \cr
h(\psi)&=&\frac{15}{8}\left(\psi-\frac{2}{3}\psi^{3}+\frac{1}{5}\psi^{
5}\right).
\end{eqnarray}
Here, $\psi$ is a nonconserved order parameter field which has one
value in the solid and another in the liquid and c is a
dimensionless diffusion field proportional to the impurity
concentration. $\Delta T=(T-T_m)/T_m$ where $T_m$ is the melting
temperature of the pure solid when $c=0$. $\Delta\psi$ is the
discontinuity in $\psi$ between solid and liquid, $K_{\psi}$ and
$K_{c}$ are phenomenological constants accounting for the free
energy cost of spatial variations in $\psi$ and $c$. The Kobayashi
function, $h(\psi)$ \cite{kobayashi93}, is introduced for analytic
and computational convenience to simplify the problem, despite the
apparent additional complexity \cite{karma98,egpk01}. It has
essentially the same effect as $\psi$ itself as $h(\psi)$ is
chosen to have the properties (i) $h(\pm 1)=\pm 1$ and (ii)
$\delta h/\delta\psi =0$ when $\psi=\pm 1$. This choice ensures
that the equilibrium values of $\psi$ are $\psi_{eq}=+1$ for
$\Delta T>0$ and $\psi_{eq}= -1$ for $\Delta T<0$ {\it
independently} of the values of $\Delta T$ and $c$ so that
$\Delta\psi=2$. The interface between the ordered $\Delta T <0$
and the disordered $\Delta T > 0$ phases is the region over which
the fields $\psi$ and $c$ change from one bulk value to the other
and has a width $\xi ={\cal O}(1)$, we define its position $(x,z)$
by $\psi(x,z)=0$.
\par To impose the motion of the interface, we take $\Delta T({\bf
r},t)=\Delta T(z-vt)$ where ${\bf v}=v{\bf\hat{z}}$ is the externally
imposed pulling velocity. Stationary states can exist  only in the
frame moving with the interface when $\Delta T({\bf r},t)=\Delta
T(z)$. In this frame, the simplest  possible dynamics consistent with
the macroscopic conservation law for $c$ are Langevin equations
 \cite{hohenberghalperin}
\begin{eqnarray}
\label{eq:dynamics}
\dot\psi=\frac{\partial\psi}{\partial t}&=&
 -\Gamma_{\psi}\frac{\delta{\cal F}}{\delta\psi}+v\frac{\partial\psi}
{\partial z}+\eta_{\psi}\equiv{\cal G}_{\psi}+\eta_{\psi} \cr
\dot c=\frac{\partial c}{\partial t}&=&\Gamma_{c}{\bf
 \nabla}^{2}\frac{\delta{\cal F}}{\delta c}+v\frac{\partial c}
{\partial z} +\eta_{c}\equiv{\cal G}_{c}+\eta_{c}
\end{eqnarray}

\noindent where $\Gamma_{\psi}$ and $\Gamma_{c}$ are the mobilities
of the fields $\psi$ and $c$ respectively. Complications such as fluid
flow are not considered. Note that the non dissipative terms
$v\partial\psi/\partial z$ and $v\partial c/\partial z$ cannot be
written as derivatives of a potential so that the dynamics of Eq.
(\ref{eq:dynamics}) is non potential. This makes the construction of a
differentiable Lyapunov functional for such situations difficult but
such a procedure has been implemented successfully in a similar
situation \cite{graham90,montagne96}. The stochastic noises
$\eta_{\psi}({\bf r},t)$ and  $\eta_{c}({\bf r},t)$ have probability
distributions
\begin{eqnarray}
\label{eq:probdistn}
 {\cal P}(\eta_{c})&&\propto{\rm exp}\Bigl[-\frac{1}{4\Gamma'_{c}}\int
 dtd{\bf r}d{\bf
r'}\eta_{c}({\bf r},t) G({\bf r},{\bf r'})\eta_{c}({\bf r'},t)\Bigr]
 \cr
{\cal P}(\eta_{\psi})&&\propto{\rm
 exp}\Bigl[-\frac{1}{4\Gamma'_{\psi}}\int dtd{\bf r}
(\eta_{\psi}({\bf r},t))^{2}\Bigr]
\end{eqnarray}

\noindent When $v=0$ in Eq. (\ref{eq:dynamics}) the system
approaches its stationary equilibrium state given by the minimum of
${\cal F}(c, \psi)$ \cite{graham90,hohenberghalperin}. $G({\bf
r},{\bf r'})$ obeys $\nabla^{2}G({\bf r},{\bf r'})=-\delta({\bf
r}-{\bf r'})$ with appropriate boundary conditions and $G^{-1}({\bf
r},{\bf r'})=-\nabla^{2}\delta({\bf r}-{\bf r'})$ so
\begin{eqnarray}
\label{eq:variance}
\langle\eta_{\psi}({\bf r},t)\eta_{\psi}({\bf
r'},t')\rangle &=& 2\Gamma'_{\psi}\delta({\bf r}-{\bf r'})
\delta(t-t') \cr \langle\eta_{c}({\bf r},t)\eta_{c}({\bf
r'},t')\rangle &=& 2\Gamma'_{c}G^{-1}({\bf r},{\bf r'})
\delta(t-t').
\end{eqnarray}

\noindent The form of Eq. (\ref{eq:variance}) is required by the
dynamics of the nonconserved field $\psi$ and the conserved field
$c$ \cite{hohenberghalperin}. If $\Gamma'_{\psi ,c} = \Gamma_{\psi
,c}$, then $\eta_{\psi}$ and $\eta_{c}$ obey a fluctuation
dissipation theorem but this is not necessary and makes no
qualitative difference. In our numerical simulations, the noise
 strengths
$\Gamma'_{\psi ,c}$ are determined by the widths of the noise
distributions. The exact form of the Green's function $G({\bf
r},{\bf r'})$ in Eq. (\ref{eq:probdistn}) is needed for numerical
computations.

\section{Path integral and state selection}

\par The model defined by Eqs. (\ref{eq:dynamics}) and
(\ref{eq:probdistn}) describes the equilibrium of the two phase
system and the invasion of one phase by another in as simple a way
as possible. Consequently, we take the noise strengths and
mobilities $\Gamma_{c,\psi}$ to be field independent constants. This
simplified model does describe equilibrium and near equilibrium
dynamics correctly and is a good phenomenological model far from
equilibrium. In our model, the temperature gradient $\Delta T({\bf
r},t)$ is taken to not affect the noise strengths. This may not be
entirely realistic but is sufficiently simple to permit analysis of
the model and does describe the essential features of directional
solidification. The great advantage of our model is that
complications such as Landauer's ``blowtorch theorem''
\cite{landauer} are not applicable despite the presence of a
``temperature'' gradient $\Delta T$ as the noise strength is taken
to be independent of $T$.
\par We formulate the problem in terms of an action ${\cal S}(\psi,
 c)$ whose minimum gives the most probable or selected state. The
joint probability distribution ${\cal P}(\psi ,c)$ is

\begin{eqnarray}
\label{eq:Z}
 {\cal P}(\psi ,c)&=&\int{\cal D}\eta_{\psi}{\cal
D}\eta_{c}{\cal P}(\eta_{\psi}){\cal
P}(\eta_{c})\delta(\eta_{\psi}-\dot\psi +{\cal
G}_{\psi})\delta(\eta_{c}-\dot{c} + {\cal G}_{c})  \cr
&=&{\rm exp}\Bigl[-{\cal S}(\psi ,c)\Bigr]
\end{eqnarray}

\noindent using Eqs. (\ref{eq:dynamics},\ref{eq:probdistn}).
The configuration $(\psi ,c)$ minimizing ${\cal S}(\psi ,c)$ is the
most probable state at least for weak noise
 \cite{dykman:01,dykman:04}.
\par The explicit expression for the action is
\begin{eqnarray}
\label{eq:action1}
{\cal S}(\psi ,c)=&&\frac{1}{4\Gamma'_{\psi}}\int  d{\bf
 r}dt\Bigl[\dot\psi({\bf r},t)
-{\cal G}_{\psi}({\bf r},t)\Bigr]^{2}  \cr
+&&\frac{1}{4\Gamma'_{c}}\int dtd{\bf r}d{\bf r'}
\Bigl[\dot c({\bf r},t)-{\cal G}_{c}({\bf r},t)\Bigr]
G({\bf r},{\bf r'})\Bigl[\dot c({\bf r'},t)- {\cal G}_{c}({\bf
r'},t)\Bigr]
\end{eqnarray}
\noindent where,  ${\cal G}_{\psi}$ and ${\cal G}_{c}$ are defined in
Eq. (\ref{eq:dynamics}) and $G({\bf r},{\bf r'})$ is the Green's
function. Note that this quadratic form for ${\cal S}(\psi ,c)$ is
very similar to the actions proposed previously
\cite{graham90,kurtze96,mardar95,bertini01}. The action of Eq.
(\ref{eq:action1}) can be obtained from the Martin-Siggia-Rose (MSR)
formalism \cite{msr73,dominicis} by integrating out the auxiliary
fields $\tilde\psi$ and $\tilde{c}$. Since the MSR action contains
all physical response and correlation functions, including those at
equilibrium. The resulting action of Eq. (\ref{eq:action1}) is also
a complete description of the dynamics and of an equilibrium system
when ${\cal G}_{\psi}=-\delta{\cal F}/\delta\psi$ and similarly for
${\cal G}_{c}$ which ensure that the dynamics of Eq.
(\ref{eq:dynamics}) is satisfied. We are interested in stationary
patterns at very large times $t$. A reasonable assumption is to
argue that the pattern becomes very close to the selected stationary
pattern at some large but finite $t_{0}$ and remains in this
stationary configuration $\psi({\bf r}), c({\bf r})$ for $t>t_{0}$
when $\dot\psi=0=\dot c$. Then, the action is dominated by the
infinitely long time interval ${\cal T}=t-t_{0}\rightarrow\infty$
and becomes
\begin{eqnarray}
\label{eq:conjecture}
\frac{4{\cal S}(\psi ,c)}{\cal
T}=\frac{1}{\Gamma'_{c}}\int d{\bf r}d{\bf r'}{\cal G}_{c}({\bf
r})G({\bf r},{\bf r'}) {\cal G}_{c}({\bf r'})
+\frac{1}{\Gamma'_{\psi}}\int d{\bf r}\Bigl[{\cal G}_{\psi}({\bf
r})\Bigr]^{2}+{\cal O}({\cal T}^{-1}), \label{eq:action2}
\end{eqnarray}
\noindent the above Eq.
(\ref{eq:conjecture}) is our essential result which should be
regarded as a conjecture as a real proof eludes us.

\par As discussed below Eq.~(\ref{eq:Z}), the configuration
$\psi({\bf r})$, $c({\bf r})$ which minimizes the action ${\cal
S}(\psi ,c)$ in Eq.~(\ref{eq:action2}) maximizes the probability
${\cal P}(\psi ,c)$. This is the selected stationary state if
dynamical paths exist between any intermediate state and the
selected state in analogy with the ergodic theorem. This is almost
certainly true but is very hard to verify by simulations as
metastable states in the Eckhaus band \cite{eckhaus} have very long
lifetimes with weak noise even for our very small systems. This is
similar to metastable states separated from the equilibrium state by
large free energy barriers. Such effects lead to apparent history
dependence of the final state in experiments \cite{trivedi} and in
theoretical investigations without noise \cite{warrenlanger}. The
essential difference in the case studied here is that the dynamics
of Eq.~({\ref{eq:dynamics}) is non potential while the approach to
equilibrium is governed by potential dynamics \cite{garcia93}. A
naive minimization of the action ${\cal S}(\psi ,c)$ of
Eq.~(\ref{eq:action1}) yields the deterministic evolution equations
which are incorrect with noise. The minimization must be done subject
to the constraints that $\psi({\bf r},t)$ and $c({\bf r},t)$ are
solutions of Eq.~(\ref{eq:dynamics}) which we perform numerically. In
principle, this can be done analytically in the MSR formalism
\cite{msr73}, as used in studies of the KPZ system \cite{fogedby99}.

\section{Numerical procedure}

\par To test our theory numerically, we generate from simulations
of Eq.~(\ref{eq:dynamics}) all quasi-stationary patterns
$\psi({\bf r}), c({\bf r})$ with noise and finite $v$ and compute
$4\Gamma'{\cal S}(\psi ,c)/{\cal T}$ from Eq.~(\ref{eq:action2}).
In the simulations, parameters similar to those describing
experiments on the liquid crystal system 4-n-octylcyanobiphenyl
(8CB) \cite{flesselles91,simon88} are used:
$\Gamma_{c}$=2$\Gamma_{\psi}$, $\Gamma_{\psi}=1$, $\gamma = 0.73$,
$K_{\psi}=1.5$, $K_{c}=0$ and the pulling speed $v=0.20$. We take
$\Delta T(z)=-T_{0}$ for $z<-W$, $+T_{0}$ for $z>W$ and $Gz$ for
$|z|<W$ with $G=T_{0}/W$ \cite{grossmann93,drolet94,drolet00}. The
temperature $T_{0}=0.38$ so that $\Delta T(z)$ is greater than the
liquidus temperature for $z>W$ and below the solidus for $z<-W$.
The simulations are carried out with the width of the temperature
variation $W=50$, the size of the simulation box $L_{z}=150$,
lattice spacing $\Delta x = 1.0$ and timestep $\Delta t=0.10$. The
system widths are $250\le L_{x}\le 1000$ with periodic boundary
conditions $\psi(x+L_{x},z,t)= \psi(x,z,t)$ and
$c(x+L_{x},z,t)=c(x,z,t)$ which restricts the allowed values of
the ${\bf q}$ vector of a periodic variation in the interface
between the ordered and disordered phases to be $q_{x}=n/L_{x}$
with $n=1,2,\cdots L_{x}$. The Green's function $G({\bf r},{\bf
r'})$ satisfies $G(x,z;x',z')=G(x+L_{x},z;x',z')$ and
$G(x,0;x',z')=0= G(x,L_{z}+1;x',z')$ and is obtained by
numerically inverting $G({\bf p})$, the discrete Fourier transform
of $G({\bf r},{\bf r'})$. This is used to calculate $\Delta{\cal
S}\equiv 4\Gamma'({\cal S}_{n}-{\cal S}_{m}) /{\cal T}$ where the
integers $n,m=q_{x}L_{x}$ label the patterns.

\begin{figure}
\includegraphics[width=7.5cm]{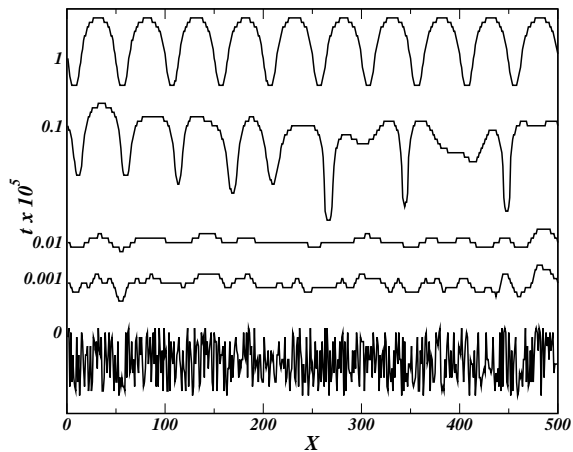}
\caption{Evolution of cellular pattern from an initial random
 interface
to final state $q_{f}=0.02$ with
velocity $v=0.20$ in absence of noise, and $L_{x}=500$.}
\label{fig2}
\end{figure}

\section{Results and Discussion}

\par We perform many simulations of Eq.~(\ref{eq:dynamics}) from
various initial conditions in order to verify if a single pattern,
out of a number of possible stationary patterns, is indeed
selected in the presence of weak noise.

First, simulations are done in the absence of noise
$\eta_{\psi}=0=\eta_{c}$ but with initial values of
$\psi=\epsilon$ in such a way to form an initial random interface
from a uniformly distributed random variable and $c=0$ for
convenience. For system widths $L_{x}=250$, $500$ and $1000$ the
system always eventually reaches a periodic stationary state with
$q_{f}=n/L_{x}=0.020$, where $n$ is the number of cells. For
$L_{x}=420$, the final periodicity is $q_{f}=8/420 \approx
0.01905$. These results imply that the selected wavevector is the
nearest allowed $q_{f}$ to the preferred value for
$L_{x}\rightarrow\infty$. In Fig. (\ref{fig2}) we show the time
evolution of the random interface for $L_x=500$, after $t=10^5$
time steps the system reaches a pattern with 10 cells, i.e,
$q_{f}=0.02$.

\begin{figure}
\includegraphics[width=7.5cm]{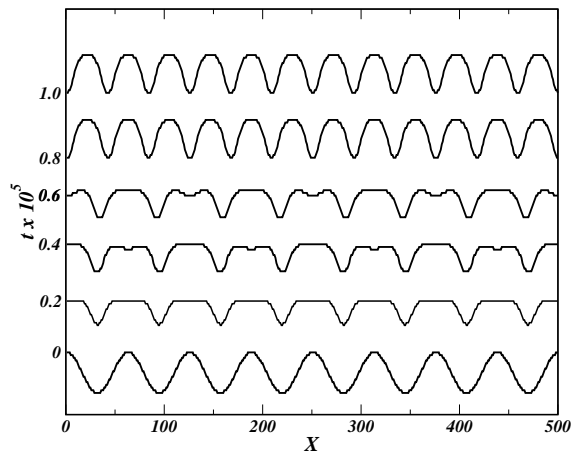}
\caption{Evolution of cellular pattern of initial periodicity
 $q_{0}=0.016$ to final state $q_{f}=0.024$ with velocity
$v=0.20$ in absence of noise, and $L_{x}=500$.}
\label{fig3}
\end{figure}

Second, we run simulations with an initial periodic interface of the
form $\zeta(x)=\zeta_{0}{\rm cos}(2\pi q_{0}x)$ with various
wavevectors $q_{0}=n_{0}/L_{x}$ and amplitude $\zeta_{0}$ similar to
that of the final state of the previous simulation, again in the
absence of noise. The initial interface is located at the center of
the simulation box of size $L_{x}\times L_{z}$. Simulations with
various allowed values of $q_{0}$ and $L_{x}=500$ yield final
stationary periodic states with $q_{f}=0.020, 0.022\,\,\, {\rm
and}\,\,\, 0.024$ which determine the Eckhaus stable band of
wavevectors for our parameter values. In an infinite system, the
Eckhaus band has a continuum of $q$ values which is reduced to a small
finite number by the finite width $L_{x}=500$. When the size is
doubled to $L_{x}=1000$, we find $7$ wavevectors in the Eckhaus stable
band from $q_{x}=0.019 \,\,\,{\rm to}\,\,\,0.025$ as expected.

\par The next simulations study the effect of noise on the
evolution of an initial periodic state with wavevector
$q_{0}=0.016$ outside the Eckhaus stable band. Without noise, the
evolution is by tip splitting of alternate cells to a final
wavevector of $q_{f}=0.024$, as shown in Fig. (\ref{fig3}). This
configuration of $12$ cells is different from the expected
wavelength of $q_{f}=0.020$ as obtained from random initial
conditions. With applied noise $\eta_{\psi}({\bf r},t)$
independently uniformly distributed with $-\eta<\eta_{\psi}({\bf
r},t)< +\eta$ with maximum noise amplitude $\eta=0.10$, the
evolution is by a totally different route to the expected state with
$q_{S}=0.020$ as shown in Fig. (\ref{fig4}). This indicates that
noise is essential in the selection of the final state and that only
one wavenumber in the Eckhaus band is truly stable in the presence
of stochastic noise. We have also performed many simulations, not
shown here because of space limitations, with several different
initial configurations all of which eventually go to the same state
with $q_{f}=0.020$ with only a single exception discussed below.

\begin{figure}
\includegraphics[width=7.5cm]{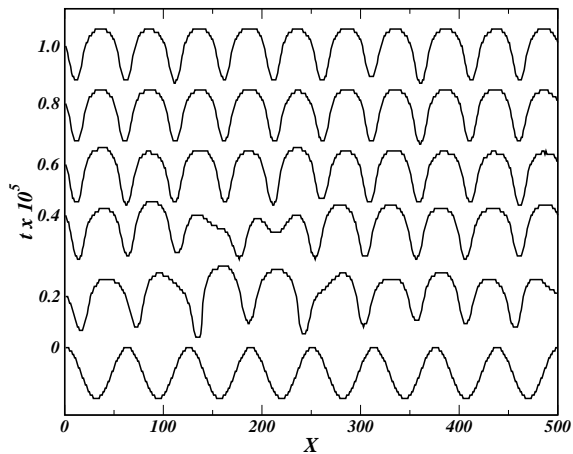}
\caption{Time evolution of same initial pattern as Fig. (\ref{fig1})
 with stochastic noise of strength $\eta=0.10$
to final state $q_{f}=0.020$. }
\label{fig4}
\end{figure}

\par Now that we have very strong evidence for selection of a
unique wavelength from almost all initial patterns except for
some $q_{0}$ in the Eckhaus band, we perform a simulation with an
initial pattern $\zeta(x)= \zeta_{0}{\rm cos}(2\pi q_{0}x)$ with
$q_{0}=0.024$ with noise to study the stability of modes in the
Eckhaus stable band \cite{CrossHohenberg,eckhaus}. In the absence of
noise, all modes with $q_{0}$ in this band are stable. We impose an
interface $\zeta(x)$, evolve it in absence of noise for $10^{4}$
time steps to allow higher harmonics to develop and then evolve the
system with noise of strength $\eta=0.20$ for $2\times 10^{5}$ time
steps with results shown in Fig. (\ref{fig6}). The interface evolves
by the gradual loss  of one cell to a final configuration of $11$
cells or $q_{f}=0.022$ which remains stable for at least $5\times
10^{7}$ steps. When a brief burst of very strong noise with
$\eta>0.25$ is applied and the system then allowed to evolve under
weak noise, the expected $q_{S}=0.020$ state results. This is
similar to the history dependent effects observed \cite{trivedi} and
discussed theoretically in the absence of noise \cite{warrenlanger}.

In order to verify our conjecture, we calculate the action from
Eq.~(\ref{eq:action2}) of the three possible states in the Eckhaus
band. As shown in Fig.(\ref{fig5}), the action difference $\Delta{\cal
S}(\psi ,c)\propto\eta^{2}$ so that, for the deterministic case,
$\Delta{\cal S}=0$. A finite $\Delta{\cal S}$ is due to finite noise
implying that the selection of a particular pattern is due to noise
driven fluctuations. Our $L_{x}=500$ system has only three
quasi-stationary periodic states in the Eckhaus stable band
\cite{eckhaus} with $q_{x}=0.020, 0.022, 0.024$ corresponding to
$n=10, 11, 12$ cells. The $q_{S}=0.020$ minimum action state is
conjectured to be the unique selected state. An outstanding question
is whether pattern selection also holds in the thermodynamic limit
$L_{x}=\infty$ when the set of stationary states in the Eckhaus stable
band is continuous. Comparing with experiment requires extending the
calculation to 3D to study the selected stationary structures of the
2D interface \cite{grossmann93,plapp04}.

\begin{figure}
\includegraphics[width=7.5cm]{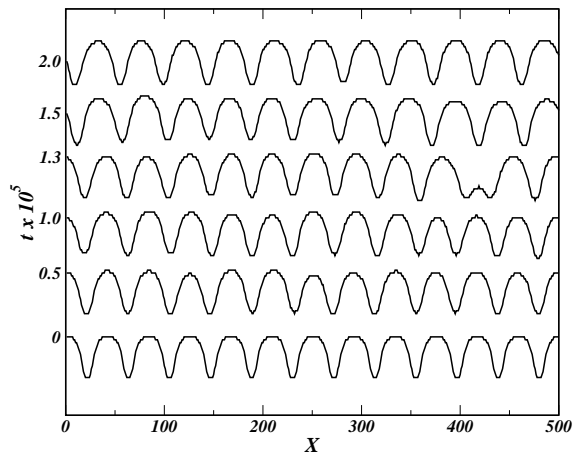}
\caption{Evolution of an interface with $q_{0}=0.024$ with noise of
 strength $\eta=0.10$ to a final state $q_{f}=0.022$. }
\label{fig5}
\end{figure}

\begin{figure}[h]
\includegraphics[width=7.5cm]{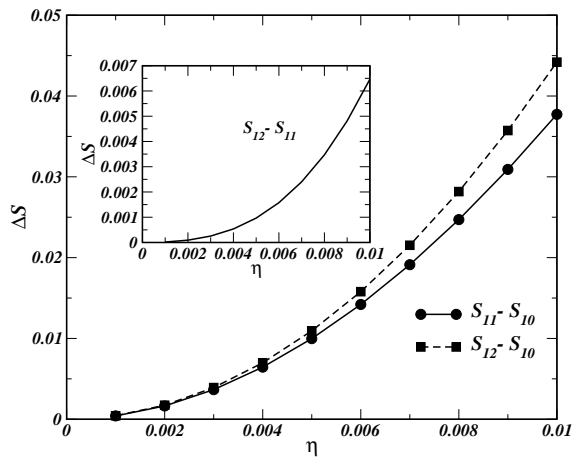}
\caption{Action differences $\Delta{\cal S} =4\Gamma'({\cal
S}_{11}-{\cal S}_{10})/{\cal T}$ and $4\Gamma'({\cal S}_{12}-{\cal
S}_{10})/{\cal T}$ from Eq.~(\ref{eq:action2}) in a system width
$L_{x}=500$ where the integers $n=q_{x}L_{x}$ in ${\cal S}_{n}$.
This shows that the pattern with wavelength $q_{S}=0.020$ has
minimum action. The noise is uniformly distributed with $-\eta
\leq\eta(\bf{r},t)\leq+\eta$.} \label{fig6}
\end{figure}

\par These numerical simulations are consistent with the hypothesis of
a unique selected stationary state with a definite periodicity
$q_{S}=0.020$. This is the noise induced stationary state  which is
reached from all initial states except from the periodic state with
$q_{0}=0.022$ in the Eckhaus  stable band. This single state is stable
with weak noise up to our longest simulation time. If a unique
selected state  exists, then only one of these should be stable to
weak noise but some are stable within our computation time and it is
impractical to perform sufficiently long runs to detect an instability
toward the  hypothetical minimum action selected state with
$q_{S}=0.020$. This study finds that all states with this single
exception do evolve to the minimum action state and, in addition,
provides some insight into the relative stability of  states in the
Eckhaus band. On balance, there is overwhelming evidence for our
hypothesis of a unique selected state  $(\psi, c)$ determined by the
minimum of the action ${\cal S}(\psi, c)$ of Eq.~(\ref{eq:action2}).
\par One important caveat must be made. In analogy with a finite
equilibrium system \cite{jljmk90}, we expect all states in our finite
system to have {\em finite} lifetimes as the stochastic noise causes
transitions between them. We expect that true stationary states exist
only in the thermodynamic limit of $L\rightarrow\infty$ when the
lifetime of the selected state should be infinite but this limit is
extremely difficult to approach numerically for well known and obvious
reasons. If a finite size scaling theory existed for these dynamical
processes, the $L\rightarrow\infty$ limit could be studied by the
methods of this letter. However, to the best of our knowledge, such a
scaling theory does not exist.

\section{Acknowledgements}

This work was supported by a NSF-CNPq grant (E.G. and J.M.K.), by
CNPq (R.N.C.F.) and FAPESP (No. 03/00541-0) (E.G.).

\end{document}